**Planar Hall effect and Anisotropic Magnetoresistance in Thin Films of Chiral Antiferromagnet $Mn_3Sn$**

Vinay Sharma, Rajeev Nepal, and Ramesh C. Budhani[*]

Department of Physics, Morgan State University, Baltimore, MD 21251 USA

[*]Corresponding author: Ramesh.budhani@morgan.edu

**ABSTRACT**

Antiferromagnetic Weyl semimetals with spin chirality offer excellent platforms to address the Berry phase physics, which manifests prominently in several of their electro-optical and electro-magnetic responses including as a large anomalous Hall effect (AHE) and spin Hall conductivity. The $Mn_3Sn$ and $Mn_3Ge$ compounds, where the Mn spins arrange in a Kagome lattice, are examples of this class of materials. Here, we report measurements of magneto-transport in c-axis textured $Mn_3Sn$ thin films grown on the [111] plane of single crystal MgO by dc magnetron sputtering. At room temperature, these films display a weak uncompensated magnetic moment of $\approx 0.12$ $\mu_B$/f.u. in the basal plane and a longitudinal resistivity ($\rho_{xx}$) close to $\approx 3.8$ $\mu\Omega$.m, which matches well with the bulk value. A residual resistivity ration ($\rho_{xx}$ (300 K)/$\rho_{xx}$ (2 K)) of $\approx 3.92$ further indicates the high quality of the films. While at 300 K a weak AHE together with field-linear Hall resistivity ($\rho_{xy}$) is observed in magnetic fields (H) applied perpendicular to the Kagome planes, the temperature (T) dependence of $\rho_{xy}$ shows prominent signatures of three magnetic phases in the temperature regime of 2 to 300 K. The $\rho_{xy}$ also derives a non-trivial topological contribution ($\rho_{THE}$ ~ 1n$\Omega$.m) in the spin glass phase which appears at T $\leq$ 100 K. The origin of the $\rho_{THE}$ is attributed to spin textures which may be appearing in a frustrated chiral spin order. Our measurements of anisotropic magnetoresistance (AMR) and planar Hall effect (PHE) over a wide H-T phase space reveal the hitherto unseen effects in the three magnetic phases of $Mn_3Sn$. While the AMR and PHE are negative in the inverse triangular spin phase (250 K $\leq$ T $\leq$ $T_N$), the helical phase (100 $\leq$ T $\leq$ 250 K) is devoid of anisotropic in-plane resistivity, and the spin glass phase shows a sign reversal of AMR with the increasing magnetic field. The origin of this sign change in AMR/PHE is attributed to the emergence of topologically protected spin textures like skyrmions where the fictitious effective magnetic field is estimated to be $\approx 4.4$ tesla.

## I. INTRODUCTION

Antiferromagnets (AFMs) are considered as the next generation spintronic materials over conventional ferromagnets (FMs) owing to their fast-switching speed, zero stray fields and stability against field–induced distortions of spin arrangements [1]. A particularly fascinating class of AFMs is presented by the compounds that lack structural inversion symmetry and possess a non-collinear arrangement of spins. Such systems host topologically non-trivial spin textures such as skyrmions[1]. The Heusler compounds of the type $Mn_3Z$ (Z = Sn and Ge) stabilized in $DO_{19}$ crystal symmetry are examples of such materials. These binary Heuslers present an excellent class of topological materials where electron correlations lead to a large anomalous Hall effect (AHE) [2], Nernst effect [3, 4], THz generated AHE [5], and other outstanding properties [5, 6]. The large spin Hall conductivity and chirality of non-collinear coplanar magnetic structure also make these compounds attractive for spintronic applications [7-10].

$Mn_3Sn$ is an AFM with the Mn spins arranged in a Kagome type inverse triangular structure stabilized due to geometric frustration and the Dzyaloshinskii-Moriya interaction (DMI) [11]. The triangular $120^0$ arrangement of Mn spins on the [0001] plane, set at the Néel Temperature of 420 K, yields a net uncompensated in-plane magnetic moment of the order $\approx$ 20 $m\mu_B$/f.u. at 300 K [2, 3]. A weak interplanar spin interaction stabilizes a helical phase at $\approx$ 250 K which changes to a spin-glass state at T < 100 K [12]. For utilizing the full functionality of $Mn_3Sn$ magnetic phases, epitaxial films of this compound have been synthesized on single crystal substrates of $Al_2O_3$, $SrTiO_3$, MgO and yttria stabilized zirconia [13-19]. The physical phenomenon that has been studied extensively in such films is the AHE [13-16]. However, a large AHE is observed only in films where the c-axis of the hexagonal structure is on the plane of the substrate. The zero-field Hall resistivity is the least in epitaxial films with out of substrate plane c-axis [14]. The large AHE and topological Hall effect (THE) in triangular AFMs have been explained by the Berry phase concept [20]. The origin of THE is attributed to the breaking of time reversal symmetry and evolution of real space Berry curvature which gives rise to a non-zero topological winding of spin textures such as skyrmions [20]. The small size and topological aspects of skyrmions may enable the potential realization of ultra-high density non-volatile memories [20, 21]. However, the non-trivial topological spin textures and skyrmions are difficult to affirmatively identify using AHE measurements. Recently, Teng Xu et al. reported the presence of Néel skyrmions in $Mn_3Sn$/CoTb interface using advanced four-dimensional Lorentz scanning transmission electron microscopy, in combination with x-ray magnetic circular dichroism photoemission electron microscopy (XMCD-PEEM) [22]. Techniques like Lorentz microscopy, XMCD-XPEEM and spin polarized scanning tunneling microscopy have been used to image spin textures in chiral AFMs [20, 23]. Interestingly, the measurements of anisotropic magnetoresistance (AMR) and planar Hall effect (PHE) provide a simpler and powerful means to address skyrmion dynamics in electrically conducting systems [24-28]. The measurements of PHE have been used by Yokouchi et. al. [24] to probe the topological spin texture in epitaxial films of MnSi. Similar measurements have been used as well to address the dynamics of spin textures in systems like $Gd_2PdSi_3$ and $Fe_5Sn_3$ [25-28]. Although the anisotropic magneto-transport (AMR/PHE) measurements have been conducted on single crystals of Mn$_3$Sn [3,29], these are limited to a small region of the H-T phase space. Moreover, to the best of our knowledge no studies of the temperature and magnetic field dependence of AMR/PHE and the signatures of skyrmion dynamics in such measurements have been reported in epitaxial films of Mn$_3$Sn.

In this work, we present an extended study in c- axis textured Mn$_3$Sn thin films grown on [111] surface of MgO. The temperature dependent magneto-transport measurements in a Hall bar geometry reveal the appearance of three magnetic phases over a temperature regime of 2 to 300 K with their distinctive signatures in AHE, AMR and PHE. The exotic topological phase in the spin glass state is also characterized by a sign reversal of AMR and PHE with the increasing magnetic field and a topological contribution to the AHE.

## II. EXPERIMENTAL DETAILS

Thin film with stack structure of [111] MgO/Mn$_3$Sn (100 nm)/Nb (2 nm) were deposited by DC magnetron sputtering of Mn$_{75}$Sn$_{25}$ (99.99%) and Nb (99.995%) targets in a load-lock multi-target high vacuum system with a base pressure of 8x10$^{-9}$ Torr. First, the MgO substrates were preheated to 700 $^0$C for 20 mins in high vacuum to remove any organic contaminants adsorbed on the surface. Since MgO is prone to adsorption of moisture, this heat treatment also allows removal of OH$^-$ radicals from its surface. After this step, the substrates were cooled to ambient temperature and Mn$_3$Sn films were deposited at the rate of 0.073 nm/s in 5 milli-Torr of ultrahigh purity (99.9999%) argon gas. The deposited films were annealed *insitu* at 500 $^0$C for 20 minutes. Finally, a thin (≈ 2 nm thick) layer of niobium was deposited on top of the Mn$_3$Sn films at room temperature for protection against oxidation. The crystallographic structure and mosaicity of the films were measured with X-ray diffraction (XRD) using a Bruker D-8 Advance X-ray diffractometer with Cu-K$_{\alpha 1}$ radiation. The Hall resistivity and anisotropic magnetoresistance of these films were measured down to 2 K in a physical property measurement system equipped with a 9-tesla superconducting magnet.

## III. RESULTS AND DISCUSSION

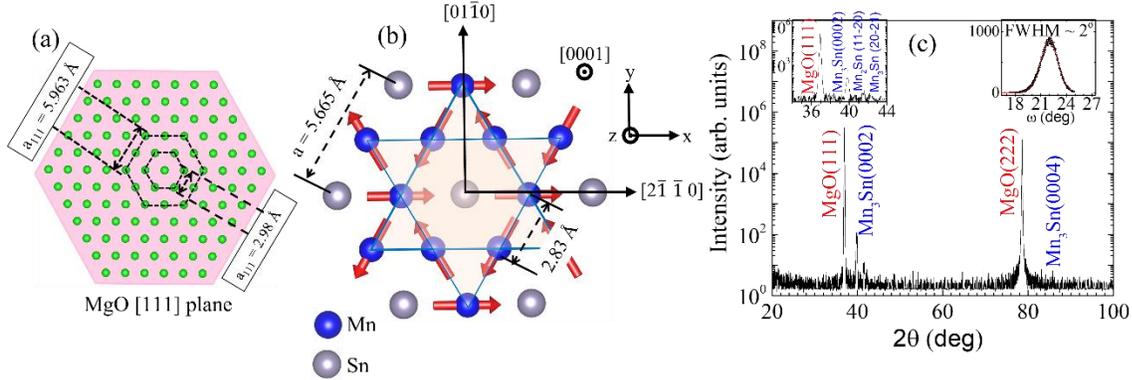

Figure 1: (a) A sketch of atomic (Mg) positions on the <111> plane of MgO, the Mg atoms are shown as green circles which form a hexagonal plane of lattice spacing 5.963 Å. (b) Atomic positions on the c-plane of the $Mn_3Sn$ lattice with in-plane ([01$\bar{1}$0], [2$\bar{1}\bar{1}$0]) and out of plane ([0001]) directions marked. The orientations of Mn spins are shown in the Kagome plane. Crystal structures are drawn using VESTA software. [30] (c) θ - 2θ X-ray diffraction pattern of a 100 nm thick $Mn_3Sn$ film grown on MgO [111] substrate. C-plane crystallographic directions are marked. Left inset shows the zoomed XRD near (0002) peak where the small peaks near $41.8^0$ and $42.2^0$ correspond to the $Mn_2Sn$ (11$\bar{2}$0) and $Mn_3Sn$ (20$\bar{2}$1) respectively. Right inset shows the rocking curve measurement near (0002) which gives full width at half maxima value of ∼ $2^0$.

$Mn_3Sn$ crystallizes in a hexagonal $DO_{19}$ type structure with space group P63/mmc [2] and lattice parameters a = 5.665 Å and c = 4.531 Å. The Mn atoms are arranged in an inverse triangular manner forming the Kagome layers stacked along the c-axis, in which the Heisenberg interaction strength on up and down triangles is different, giving rise to a 'breathing' anisotropy. The epitaxial relationship between [111] MgO and $Mn_3Sn$ is illustrated in Fig. 1(a) and (b). Figure 1(c) shows the θ-2θ XRD pattern of a 100 nm $Mn_3Sn$ thin film where the presence of [0002] and [0004] reflections of the hexagonal $Mn_3Sn$ confirms that the [000n] reflections appear next to the cubic [111] reflections of MgO [31]. $Mn_3Sn$ lattice constant c of 4.530 Å, determined from the [0002] reflection is close to the bulk value [2]. The peaks near 2θ = $41.8^0$ and $42.2^0$ in Fig. 1(c) are attributed to the $Mn_2Sn$ (11$\bar{2}$0) and $Mn_3Sn$ (20$\bar{2}$1) respectively arising presumably due to the Mn deficiency [15]. The Mn deficient samples show a helical magnetic phase transition near 250 K due to the formation of $Mn_2Sn$, which has a Curie temperature of ≈ 250 K [32]. The significance of this transition will be addressed further in the following sections.

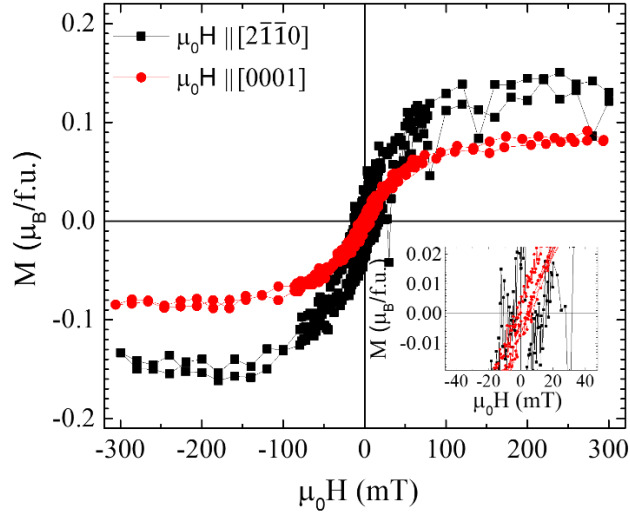

Figure 2: Magnetization of a 100 nm thick Mn$_3$Sn film at two orientations of the external field; in-plane parallel to ([2$\bar{1}\bar{1}$0]) and out of plane, parallel to ([0001]) directions. Inset shows a magnified view of data near zero field to differentiate the coercive field in the in-plane and out of plane directions.

Figure 2 shows the room temperature magnetization loops of a Mn$_3$Sn film measured using a vibrating sample magnetometer (VSM). These data have been corrected for the diamagnetic response of the MgO substrate. The sample shows a weak ferromagnetic response with a saturation moment of $\approx$ 0.12 $\mu_B$/f.u. for the in-plane field. For the out-of-plane field, the moment drops to $\approx$ 0.08 $\mu_B$/f.u. The weak moment seen for the in-plane field indicates a small canting of the Mn spins in the a-b plane. Whereas the minuscule moment in the out-of-plane direction suggests some canting along the c-axis as well. The small in-plane and out of plane moments seen here are an order of magnitude higher than the reported values for Mn$_3$Sn crystals (4-7 m$\mu_B$/f. u) [2]. This discrepancy has been discussed previously in the context of disorder and Mn deficiency in thin films, which in turn generate the out of plane tilting of Mn moments [13,33]. The in-plane uncompensated moment in these epitaxial Mn$_3$Sn films were found to be $\sim$ 0.21 $\mu_B$/f.u which is approximately double of the moments found in the present case. In these reports [13,33], the larger magnetization is attributed to the uncompensated moments arising due to the disorder and defects caused by excess of Sn atoms randomly occupying Mn sited as well as Mn vacancies [13]. We therefore argue that the Mn$_3$Sn films in present case possess lower disorder and antisite defects. We also note an anisotropy in the coercive field (H$_c$) of these films as shown in the inset of Fig. 2. The coercive field when the field is orientation along the [2$\bar{1}\bar{1}$0] direction ($\mu_0$H$_c \approx$ 10 mT) is higher by a factor of $\approx$ 3, as compared to the H$_c$ for the field orientation along [0001]. This is presumably due to the mosaicity of epitaxy in the ab-plane of the film and the measured magnetization reflects an average over many crystallites [14].

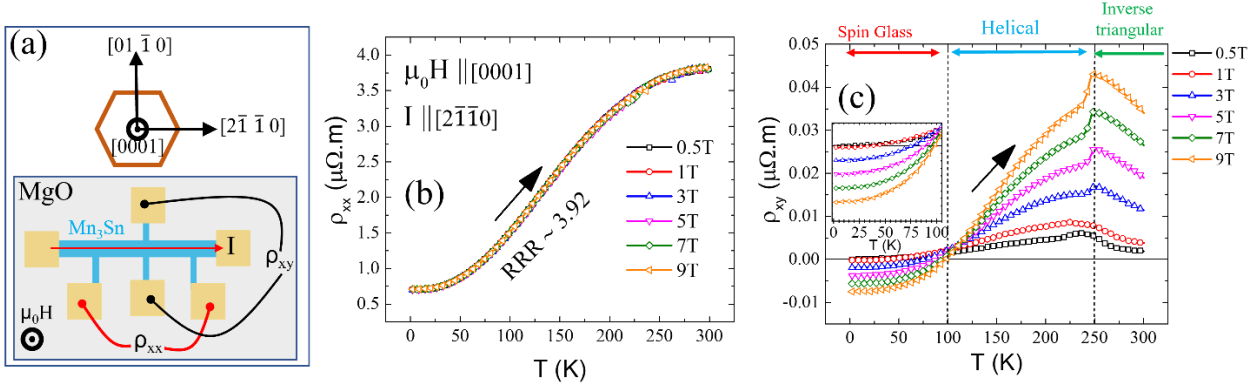

Figure 3 (a) Schematic representation of magneto-transport measurements in a Hall bar geometry. (b) Longitudinal resistivity ($\rho_{xx}$) measured as a function of temperature at several values of magnetic field applied perpendicular to the plane of the film (<0001>) during field warming. (c) Hall resistivity ($\rho_{xy}$) measured as a function of temperature at different values of the out-of-plane magnetic field. The Hall resistivity clearly shows transitions from spin glass to helical and from helical order to inverse triangular antiferromagnetic order on warming from 2 K to 300 K. These transitions are marked by dotted vertical lines. Inset shows a magnified view of the Hall data at T < 100 K.

The temperature dependent magneto-transport measurements offer a powerful means to probe magnetic phase transitions in Mn$_3$Sn thin films. Here we measure the longitudinal ($\rho_{xx}$) and Hall ($\rho_{xy}$) resistivity of a 100 nm thick Mn$_3$Sn film patterned in a Hall bar geometry, as shown in Fig. 3(a). The value of $\rho_{xx}$ is extracted after symmetrizing the +H and -H data i.e., $\rho_{xx} = (\rho_{xx}(+H) + \rho_{xx}(-H))/2$ whereas the true $\rho_{xy}$ is obtained on antisymmetrization of the transverse resistivities i.e., $\rho_{xy} = (\rho_{xy}(+H) - \rho_{xy}(-H))/2$. Figure 3(b) shows the $\rho_{xx}$ (T) data, which confirms the metallic nature of Mn$_3$Sn as reported in works on bulk crystals [2]. The room temperature $\rho_{xx}$ of 3.8 µΩ.m is close to the reported value (3.1µΩ.m) in epitaxial thin films [14]. The residual resistivity ratio (RRR ~ 3.92) marked in Fig. 3(b) is higher by a factor of ≈ 3 as compared to the result of Taylor et.al. [14], suggesting a better-quality film. Fig. 3(b) also shows that the $\rho_{xx}$ curves measured at different magnetic fields overlap, indicating a very low (≈ 0.1 – 0.3%) magnetoresistance for the out of film plane oriented magnetic field. This value is comparable to the previously measured MR in out of plane field [14] and indicates the presence of weak magnetization in these AFM films.

The non-collinear AFMs like Mn$_3$Sn are known to display a giant AHE which could be used to identify different magnetic phase transitions in the material [3]. We have measured the $\rho_{xy}$ of the Mn$_3$Sn films over a broad range of temperature and magnetic field strengths. Figure 3(c) shows the temperature dependent $\rho_{xy}$ measured with the magnetic field applied along the [0001] direction (perpendicular film plane). These data reveal two critical temperatures where the $\rho_{xy}$ undergoes a dramatic change. First a well-defined cusp develops at T ≈ 250 K, which agrees well with the reported temperature of a magnetic transition from the inverse triangular spin structure to a helical phase on cooling below ambient temperature. The sharp drop in $\rho_{xy}$ at this temperature may be related to a change in the mechanism of scattering or a change in the Berry curvature of electronic states at the Fermi energy. The second critical behavior of $\rho_{xy}$ emerges at T ≈100 K where the system enters in a spin glass state on cooling [34, 35]. The $\rho_{xy}$ below this critical point undergoes a sign change, which is evident in the magnified plot of the data at T < 100 K shown in the inset of Fig. 3(c). While a giant zero-field AHE is seen in Mn$_3$Sn crystals when measured with field in the a-b plane as a result of the non-collinearity of spins in the Kagome planes [2,14], the out of ab-plane field data presented in Fig. 3(c) is suggestive of spin canting along the c-axis in the presence of a strong field [14]. We tentatively attribute the change in sign of the $\rho_{xy}$ below ≈100 K to the emergence of spin textures in the spin glass phase which may contribute to a THE that counteracts the spin canting contribution to $\rho_{xy}$ above 100 K. We discuss this feature of $\rho_{xy}$ in the next section.

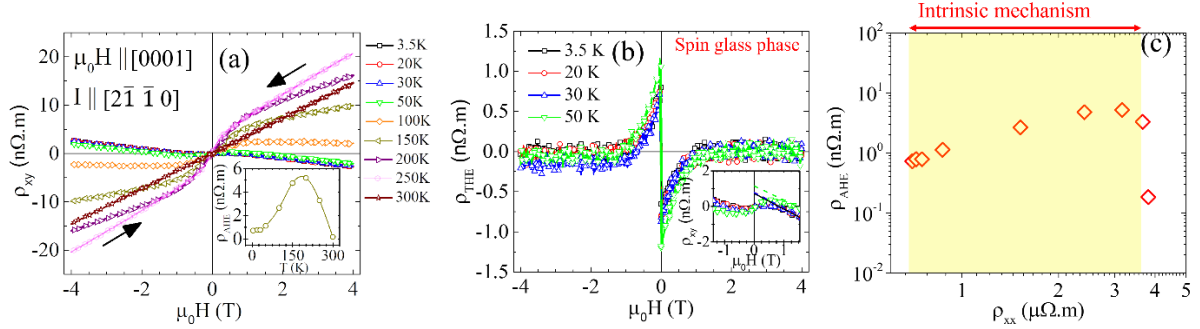

Figure 4: (a) Hall resistivity at several temperatures plotted as a function of magnetic field applied parallel (//) to the film normal (// c-axis) with current flow along [01$\bar{1}$0] direction. Inset shows the Hall resistivity derived from AHE contribution ($\rho_{AHE}$) which is calculated by the linear fitting at high field and extrapolated to the zero-field value. (b) Derived topological Hall effect in the spin glass phase (T< 100 K). The inset of 4(b) shows the zero-field extrapolation of the Hall data to extract the topological contribution (see text). (c) Scaling relation between $\rho_{AHE}$ and $\rho_{xx}$ at different temperatures shows that Mn$_3$Sn sits in a good metal regime where AHE generates due to intrinsic mechanism.

We now present the measurements of Hall resistivity as a function of applied magnetic field at several temperatures to address its hysteretic behavior and its value at zero field. While the former is important to establish the orientational dynamics of Mn moments, the latter is related to the momentum space fictitious field seen by the conduction electrons. Figure 4(a) shows the variation of $\rho_{xy}$ as a function of the out- of- plane magnetic field when current (I) is applied parallel to [2$\bar{1}\bar{1}$0] direction. A linear Hall response ($d\rho_{xy}/dH$ = constant) is observed at 300 K. This linearity is primarily due to the ordinary Hall effect ($R_0 \cdot \mu_0 H$) because the weak magnetization of Mn$_3$Sn lies along the basal plane (a-b plane) which is orthogonal to the applied field direction. The carrier density obtained from this linear Hall response using the Drude relation is ≈ 1.70 x $10^{21}$ cm$^{-3}$. In the temperature window of 250 K to 100 K we see a clear non-linear Hall response in field $\mu_0 H \leq$ 1 tesla. The linear extrapolations of the high field $\rho_{xy}$ ($\pm\mu_0 H$) data to zero field and antisymmetrization of these zero-field values yield a zero-field AHE, which peaks to ≈ 5 nΩ.m at 200 K. It is instructive to compare this number with the value of zero-field $\rho_{xy}$ in polycrystalline and [0001] oriented epitaxial films of Mn$_3$Sn. Previous report on epitaxial thin films shows a very low value of $\rho_{AHE}$ ($\rho_{AHE}$ <1 nΩ.m) in c-plane and significantly higher value ($\rho_{AHE}$ > 2 nΩ.m) a-b plane-oriented samples [14]. In polycrystalline Mn₃Sn thin films, negative AHE is observed with a large coercivity [36].

A strikingly different Hall response is seen at T ≤ 100 K. The zero-field $\rho_{xy}$ is now very small, and at higher fields it is linear with a negative slope. However, the zero-field value of $\rho_{xy}$ remains positive albeit very small, as shown in the inset of Fig. 4(b). Since Mn₃Sn is known to transit to a spin-glass phase at T< 100 K, the frustrated Mn moments in this temperature regime may form topologically non-trivial spin textures which may lead to a topological contribution to $\rho_{xy}$. Consequently, the net Hall resistivity of our chiral magnet at T < 100 K may be expressed as [37],

$$\rho_{xy} = R_0\mu_0 H + R_S\mu_0 M_Z + \rho_{THE} \quad (1),$$

where the first term accounts for the ordinary Hall effect and the second term is the anomalous Hall which derives contributions from scattering and band structure effects such as anomalous velocity and Berry curvature [38]. The last term in Eq. 1 is the the topological Hall resistivity ($\rho_{THE}$). The value of $\rho_{THE}$ is calculated after subtracting the contributions of ordinary Hall and AHE from the measured $\rho_{xy}$. In Fig.4(b) we show the $\rho_{THE}$ at T < 100 K as a function of the out- of- plane magnetic field. To find the $\rho_{THE}$, we first calculate the ordinary Hall contribution from the slope of the field linear Hall data. The AHE contribution is calculated by extrapolating the linear response to the zero field. The value of $\rho_{THE}$ (H) is then extracted by subtracting the two contributions from measured $\rho_{xy}$ (H). Figure 4(c) shows the relation between AHE resistivity and longitudinal resistivity to justify the standard scaling law of "good metal" regime. The relation $\rho_{xy} \approx \rho_{xx}^\alpha$ with α = 1 suggests that the AHE in Mn₃Sn is governed by the induced Berry curvature [38].

Electronic band structure calculations show that Mn₃Sn is a magnetic Weyl semimetal [39]. Together with the observation of a large zero-field anomalous Hall resistivity, the presence of a strong anisotropic backscattering of charge carriers in this material is an indicator of a time reversal symmetry broken Weyl semi-metallic. Here, we report, for the first time, the behavior of AMR and PHE in the Mn₃Sn epitaxial films whose Hall and out – of - plane magnetoresistance data are reported in the earlier section.

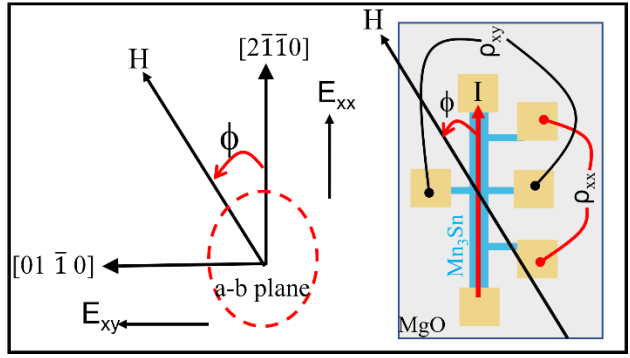

Figure 5 Sketch of Longitudinal ($\rho_{xx}$) and transverse ($\rho_{xy}$) resistivity measurements as a function of angle ($\Phi$). Here, current I is flowing along [2$\bar{1}\bar{1}$0] and magnetic field is applied in a-b plane. $\rho_{xx}$ and $\rho_{xy}$ are measured along [2$\bar{1}\bar{1}$0] and [01$\bar{1}$0] directions respectively. Corresponding electric fields $E_{xx}$ and $E_{xy}$ are also shown in this figure.

Unlike the AHE, AMR and PHE are even magneto-transport effects, and thus, symmetric under magnetic field reversal. Figure 5 shows a schematic of the AMR/PHE measurement geometry where current and magnetic field are applied in the plane of the film while it is given an in-plane rotation of 360 degrees. This allows a continuous variation of the angle $\Phi$ between the current and magnetic field.

In a typical 3d ferromagnet like permalloy, the $\phi$ dependence of $\rho_{xx}$ and $\rho_{xy}$ for in-plane field is expressed as [40-42]

$$\rho_{xx}(\Phi) = \rho_\perp + (\rho_\parallel - \rho_\perp)\cdot \cos^2(\Phi) \qquad (2)$$

$$\rho_{xy}(\Phi) = (\rho_\parallel - \rho_\perp)\cdot \cos(\Phi)\cdot \sin(\Phi) \qquad (3),$$

and the difference ($\rho_\parallel - \rho_\perp$) defines the $\rho_{AMR}$. The physical origin of AMR has been attributed to anisotropic *s-d* backscattering of conduction electrons due to high spin orbit coupling of 3d orbitals [40, 41]. This leads to a difference in the resistivity ($\rho_\parallel$) when the current (I) and magnetization (M) are parallel and the resistivity ($\rho_\perp$) when the I and M are orthogonal. While the difference ($\rho_\parallel - \rho_\perp$) is generally positive in 3d ferromagnets, this may not be the case in topological insulators and Weyl semimetals due to their special band topology, that leads to an interband contribution to conductivity and a non-zero Berry curvature of states near the Fermi energy which may be connected to chiral anomaly in transport [40].

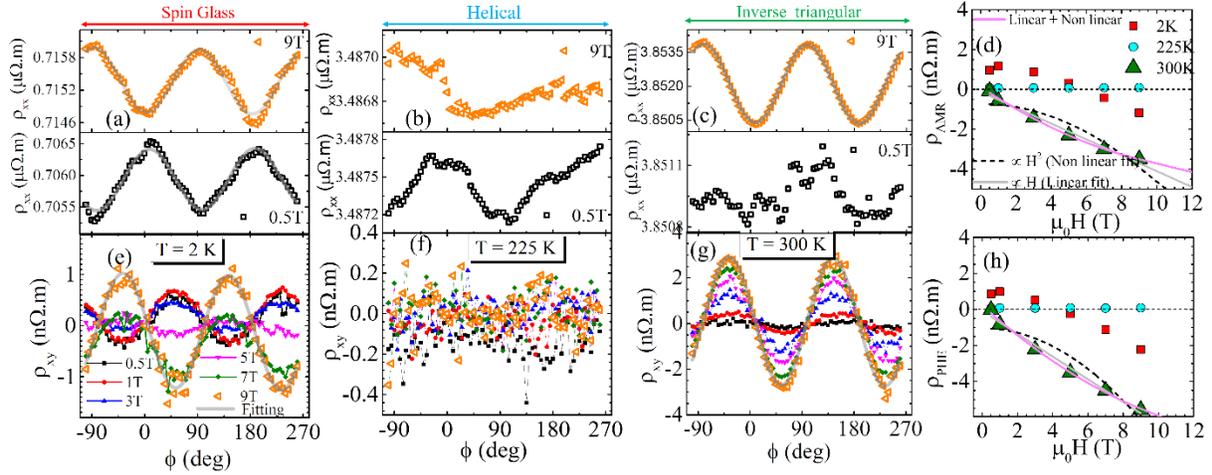

Figure 6 (a-c): Longitudinal resistivity ($\rho_{xx}$) as a function of angle Φ measured in three different magnetic phases (T < 100 K, 100 K < T < 250 K and T > 250 K) of $Mn_3Sn$ at lower (0.5 T) and higher (9 T) magnetic fields. (d) The calculated AMR as a function of magnetic field in the three magnetic phases. (e-g) Transverse resistivity ($\rho_{xy}$) as a function of angle Φ measured in the temperature regimes T< 100 K, 100 K < T < 250 K and T > 250 K) corresponding to the spin glass, chiral and    phases, respectively. (h) Derived planar Hall resistivity as a function of magnetic field in three different magnetic phases of the film. The data taken in the inverse triangular phase (T > 250 K) have been fitted to a particular type of H- dependence. Solid gray line and magenta line fitting in (d) and (h) corresponds to linear and second order polynomial relation while broken black line corresponds to quadratic ($H^2$) fitting. The $H^2$ dependence corresponds to the 3d ferromagnets while the combination of linear and quadratic fitting is the signature of chiral anomaly.

We have measured the Φ-dependence of $\rho_{xx}$ and PHE of the $Mn_3Sn$ films. Figure 6(a) shows the variation of $\rho_{xx}$ at 0.5 and 9.0 tesla as a function of angle Φ measured at 2 K, where the $Mn_3Sn$ is in the spin glass phase. Here we emphasize that these data have been obtained after symmetrizing the measurements performed at the negative and positive field orientations to eliminate any contribution of anomalous Hall voltage that might creep in due to a misalignment between the magnetic field and the plane of the sample. While these data broadly follow the angular dependence predicted by Eq. 2, a sign reversal of AMR with increase in magnetic field at T < 100 K suggests a magnetic field induced change in charge career scattering. This field induced change in sign of ($\rho_\parallel - \rho_\perp$) is also seen in the $\rho_{xy}$ data of the spin glass phase (see Fig. 6e).

In Fig. 6(b) and Fig. 6(f) we show the angular dependence of $\rho_{xx}$ and $\rho_{xy}$ at 225 K. Barring a small variation in $\rho_{xx}$ at 0.5 tesla, which is not in agreement with Eq. 2, the AMR and PHE display no angular dependence in the helical phase. This observation seems to suggest that the anisotropic backscattering here is directly related with the magnetization, which is close to zero in the helical phase [2]. The AMR reappears along with the PHE when the sample temperature is raised above 250 K, where it is in the inverse triangular phase, as seen in Fig. 6(c) and Fig. 6(g) respectively. However, unlike the case of the spin glass state, the AMR and PHE remain negative over the entire field range. Figure 6 (d) and (h) summarize the behavior of $\rho_{AMR}$ and $\rho_{PHE}$ as a function of magnetic field in the three magnetic phases. In the spin glass phase (T = 2 K), the sign change in AMR and PHE from positive to negative occurs on increasing the field beyond ≈ 4 tesla. We have analyzed the field- dependence of the AMR and PHE measured at 300 K in the framework of a second order polynomial and a quadratic function of magnetic field. A positive AMR with quadratic field dependence is generally seen in itinerant ferromagnets due to the classical anisotropic orbital magnetoresistance [43]. The data of Fig. 6(d &h) clearly deviates from this behavior and shows a combination of linear and quadratic dependencies with negative overall values which may be attributed to chiral anomaly in the Weyl state. This type of higher order harmonics in AMR/PHE can be a signature of Weyl fermions in $Mn_3Sn$, as previously reported in Kagome type AFMs [43]. The sign change in the 2 K data presumably reflects a crossover from the classical regime to the chirality dominated transport.

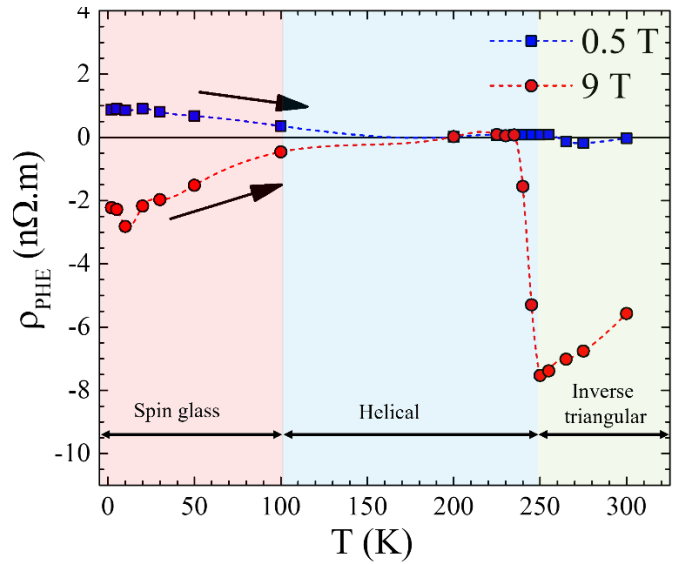

Figure 7 Temperature dependent PHE resistivity calculated by fitting the $\rho_{xy}(\phi)$ data taken at 0.5 and 9.0 tesla in-plane field to Eq. 3. All three magnetic phases of $Mn_3Sn$ are revealed clearly by the PHE data. The square and circular data points are measured values and dotted lines are guides to the eye.

The magnetic phases of $Mn_3Sn$ are emphasized better by the temperature dependence of PHE measured at 0.5 and 9.0 tesla, as shown in Fig. 7. The PHE and AMR are absent in the helical phase at all fields. The spin reorientation points at 250 and 100 K match well with the reported neutron diffraction studies [34]. The PHE signal is mainly dependent on the magnetic moment which is a continuously increasing function of applied magnetic field due to the non-collinear AFM nature of Mn$_3$Sn. At T > 250 K, Mn$_3$Sn magnetic moments lie in the a-b plane forming a co-planar inverse triangular orientation. This temperature dependent PHE behavior mimics the field cooled magnetization behavior of Mn$_3$Sn up to 100 K [12,44]. However, the increment in PHE and the peculiar sign change at low temperature (T < 100 K) shown in Fig. 7 cannot be understood only by the temperature dependent magnetization data of Mn$_3$Sn. We have done a detailed study of temperature dependent PHE in this range to shine light on this unexplored region.

A sign change in AMR/PHE driven by temperature and magnetic field has been observed earlier in various systems [40]. The sign reversal phenomenon in very thin films (~ 4 nm) of hole doped manganites on changing the magnetic field has been attributed to a change in magnetic anisotropy as the state of strain in the field changes with thickness. [45]. A temperature dependent sign change in AMR/PHE has also been observed in topological insulators [46]. Similar effects in AMR/PHE are seen as well in various Dirac and Weyl semimetals where the chiral charge pumping, and spin-flip back scattering break the time reversal symmetry [40]. However, to the best of our knowledge, no similar effects have been reported in Mn$_3$Sn thin films.

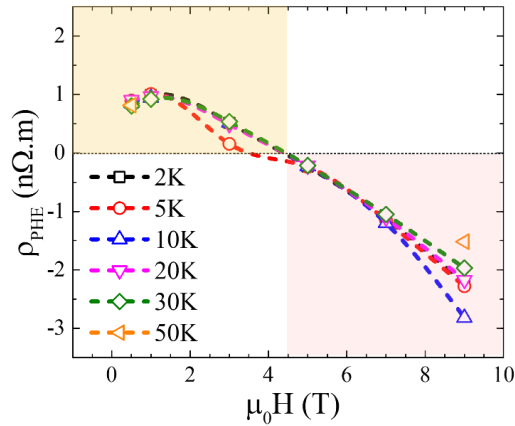

Figure 8 Magnetic field dependence of PHE at several temperatures in the spin glass state of Mn$_3$Sn. The broken lines in the figure are guides to the eye.

We now address the sign reversal of PHE in the spin glass phase in greater detail. Figure 8 shows the field dependence of the PHE at T ≤ 50 K at several temperatures. These measurements are performed in such a manner that the sample is first demagnetized by an oscillating magnetic field to remove any residual magnetization and then the field of positive and negative polarity is applied to collect the $\rho_{xy}(\phi)$ data, as depicted in Fig. 5. The PHE remains positive as long as the magnetic field is less than ≈ 4.5 tesla. At H > 4.5 tesla, its reverse sign. This crossover point remains the same (≈ 4.5 tesla) at all temperatures below ≈ 50 K. Previous low temperature neutron diffraction studies have identified a spin glass phase in $Mn_3Sn$ characterized by a slight random tilting of Mn-moments away from the c-plane [47]. Two possible scenarios have been presented for the canting on Mn-moments: (i) all three magnetic moments of the inverse triangular structure tilt toward positive c-axis and (ii) one moment tilts downwards and the other two tilt upward from the c-plane. The second scenario is presumed to lead to the non-trivial topological spin textures and consequently the topological Hall effect [20]. The strength of THE in these $Mn_3Sn$ films at T ≤ 50 K has been presented earlier in Fig. 4(b). This observation of THE in $Mn_3Sn$ is also consistent with the previous reports on non-coplanar magnetic systems like MnSi, $Gd_2PdSi_3$ and FeSb where the low temperature topological spin textures have been attributed to the skyrmionic state [48 - 50].

We now address the sign change of PHE in the spin glass state of $Mn_3Sn$ at a field of ≈ 4.5 tesla in the framework of topological spin textures by expressing the $\rho_{THE}$ as[20, 49],

$\rho_{THE} = P \cdot R_0 \cdot \mu_0 H_{eff}$  (4),

where P is the local spin polarization of conduction electrons, $R_0$ is the normal Hall coefficient and $H_{eff}$ is the fictitious effective field arising from the Berry phase which orients opposite to the applied field [20]. We may now compare the value of the magnetic field at which PHE changes sign (Fig. 8) with the calculated value of $H_{eff}$. For $Mn_3Sn$ thin film, $\rho_{THE}$ ≈ 1 nΩ.m (calculated from Fig. 4(b)), the value of $R_0$ calculated from Fig. 4(a) is ≈ 8.5 x $10^{-10}$ Ω.m/tesla and the P is defined as the ratio of ordered magnetic moments in spin glass phase to the saturated magnetic moments and taken as P ~ 0.27 which is in a good agreement to the previously calculated values [11]. After putting all these values in Eq. 4, the fictitious effective field $\mu_0 H_{eff}$ is ≈ 4.4 tesla, which is in remarkable agreement with the experimental value obtained in Fig. 8. Hence, we may argue that the non-trivial topological spin texture and PHE sign change in the spin glass phase are closely related. Further exploration of this fictitious magnetic field and skyrmions in $Mn_3Sn$ are suitable subjects for future studies.

## IV. Summary and conclusion

In summary, we have conducted detailed measurements of anisotropic in-plane magneto-transport and anomalous Hall effect in c-axis oriented thin films of chiral antiferromagnet $Mn_3Sn$ over a wide H-T phase space. These films were deposited on the [111] surface of MgO crystals using ultrahigh vacuum magnetron sputtering. The magnetic hysteresis loops of the films measured at 300 K reveal a weak magnetic moment arising due to the canting of Mn spins towards the c-axis. We have established the presence of three distinct magnetic phases in this chiral magnet through the measurements of AHE, AMR and PHE over a temperature range of 2 to 300 K. While the helical phase stabilized between $\approx 100$ to 250 K is devoid of anisotropic in-plane transport, the spin glass phase formed at T < 100 K is characterized by a magnetic field dependent sign reversal of AHE and PHE and a topological contribution to the Hall voltage which presumably comes from the non-trivial spin textures of the spin glass phase. Notably, no such sign change of AHE/PHE is seen in the inverse triangular spin state of the system which occurs above $\approx 250$ K. Direct imaging of the spin textures with Lorentz transmission electron microscopy in spin glass phase would be highly desirable direction for further studies.

## Acknowledgements

This research is funded by the Air Force Office of Scientific research under grant # FA9550-19-1-0082. We thank Dr. Ravinder Kumar for collecting the X-ray diffraction data and Dr. Rajeswari Kolagani at Towson University for providing access to the X-ray diffractometer.